\documentclass[epj]{svjour}
\usepackage{graphics}
\usepackage{graphicx}
\usepackage{dcolumn}
\usepackage{bm}
\usepackage{color}

\newcommand{\be}{\begin{equation}}
\newcommand{\ee}{\end{equation}}
\newcommand{\ba}{\begin{eqnarray}}
\newcommand{\ea}{\end{eqnarray}}
\newcommand{\bd}{\begin{displaymath}}
\newcommand{\ed}{\end{displaymath}}
\newcommand{\bea}{\begin{eqnarray}}
\newcommand{\eea}{\end{eqnarray}}

\renewcommand{\vec}[1]{\mbox{\boldmath$#1$}}

\def\thalf{{\textstyle{\frac{1}{2}}}}

\def\oneqt{{\textstyle{\frac{1}{4}}}}

\begin{document}
\title{A study of $\Lambda$ and $\bar{\Lambda}$ polarization splitting by meson field in PICR hydrodynamic model}

\author{Yilong Xie\inst{1} \and Gang Chen\inst{1} 
\and Laszlo Pal Csernai\inst{2,3}}                     
%
%
\institute{School of Mathematics and Physics, China University of Geosciences (Wuhan), 
Lumo Road 388, 430074 Wuhan, China 
\and
Department of Physics and Technology, University of Bergen,
Allegaten 55, 5007 Bergen, Norway
\and
Frankfurt Institute for Advanced Studies, Ruth-Moufang-Strasse 1,
60438 Frankfurt am Main, Germany
}
\date{Received: date / Revised version: date}
%
\abstract{
With the PICR hydrodynamic model, we study the polarization splitting between $\Lambda$ and $\bar{\Lambda}$ at RHIC BES energy range, based on the meson field mechanism. Our results fit to the experimental data fairly well. 
Besides, two unexpected effect emerges: (1) the baryon density gradient has non-trivial and negative contribution to the polarization splitting; (2) for 7.7 GeV Au+Au collisions within the centrality range of 20\%-50\%, the polarization splitting surprisingly increases with the centrality decreases. The second effect might help to explain the significant signal of polarization splitting measured in STAR's Au+Au 7.7 Gev collisions.
} 
\maketitle
\section{Introduction}

Non-central heavy ion collisions create a participant system of extremely hot and dense matter, carrying substantial angular momentum that is perpendicular to the  reaction plane 
\cite{BPR2008,GCD2008,VAC2014}. 
Through the spin-orbital coupling, just as the Einstein-de-Hass effect 
\cite{EH1915} and Barnet effect 
\cite{B1935} 
had revealed, the initial fireball angular momentum will eventually give rise to the spin alignment of final particles, such as $\Lambda$ hyperons
\cite{BCDG13,BCW2013}. 
The $\Lambda$ hyperon reveals its polarization by emitting preferentially the weak decay products along its spin direction, and thus is a fairly good choice of polarization measurement in experiments 
\cite{IA06,IS06,BIA07}.  
Many theories and simulations were also addressing this topic 
\cite{BCDG13,LW05a,HHW11,BGT07}.

Recently, the STAR collaboration measured the non-vanishing $\Lambda$ polarization for Au-Au collisions  at different energies 
$\sqrt{s_{\rm NN}}=$ 7.7 - 200 GeV
\cite{FirstSTAR,Nature,SecondSTAR}, 
and as far as we know, the results conform with the theoretical predictions and simulations in two significant aspects: the global polarization of both $\Lambda$ and $\bar{\Lambda}$ aligns with the  initial angular momentum, and decreases with the energy; the local polarization along the beam direction shows quadrupolar structure on transverse momentum plane. 

However, there still exist some puzzles in this field \cite{QMSTAR}. 

Locally, the longitudinal polarization on transverse momentum plane, from model simulations of both a multiple phase transport (AMPT) model  
\cite{XLTW2018} 
and the hydrodynamic model \cite{Xie2016,BK2018}, exhibits opposite signature to the experimentally observed quadrupolar structure\cite{STAR2019}. Many recent works have been devoted to this problem\cite{SK2019,WPHW2019,FKMR2019}, and the feed-down effect from hyperon decay was proved to be too trivial to explain\cite{XLHH2019,BCS2019}.Our recent work 
\cite{XWC2019} 
using the high resolution (3+1)D Particle-In-Cell Relativistic (PICR) hydrodynamic model to calculate the polarization at 200 GeV Au-Au, shows a fairly good agreement to the experimentally observed longitudinal polarization.

Globally, the magnitude of $\bar{\Lambda}$ polarization is larger than that of $\Lambda$ polarization. Some might argue that due to the large errors in measurements, it is not sure that whether this polarization splitting really exists, but at least for collision energy of $\sqrt{s_{\rm NN}}=$ 7.7 GeV, the splitting effect  can be identified with high confidence level (see Fig. 4). This splitting effect has raised great interests. It was proposed that the magnetic field induced by the charged spectators can give rise to the polarization splitting between $\Lambda$ and $\bar{\Lambda}$, but this will require a magnetic field that is long lasting and has a large magnitude, which are not very realistic. Besides, as indicated by Relativistic Magneto-hydrodynamics, the magnetic field can also be induced by charged particles in vortical Quark-Gluon-Plasma (QGP), and in this scenario the magnetic field could last long enough until freeze-out, but problem still exists: the charge density might not be large enough to produce a magnetic field that is strong enough. E.g., the upper limits of the estimated polarization difference at 7.7 GeV is below 1\%, which is far away from the lower boundary of experimentally observed 3\% difference \cite{GLW2019} . 

Another novel mechanism was proposed by Ref. \cite{CKW2019}, that the vector meson's "magnetic" field, induced by the baryon vorticity at freeze-out, can split the polarization. However, the polarization splitting formula therein is driven mainly by the directed flow coefficient ($c_1$) and the shear flow coefficient ($c_3$)
\cite{CKW2019}.
The coefficient $C$, which is proportional to $\Delta c = c_1 - c_3$, is actually a free parameter. 

Therefore, in this paper, we are going to revisit the theory in Ref.  \cite{CKW2019} and modify the polarization splitting formula therein, by removing the free parameter $C$ and explicitly bringing out the vorticity, which is essential in $\Lambda$ polarization study. Then based on this meson field mechanism, we apply the PICR hydrodynamic model that has been previously used in polarization studies, to simulate and calculate the polarization splitting effect.

The paper is organized as follows. In Section 2, the theory in Ref. \cite{CKW2019} is revisited and modified to explicitly include the vorticity and baryon density gradient. In section 3, we use the PICR hydrodynamic model to simulate and calculate the polarization splitting for Au+Au collisions at RHIC BES energies $\sqrt{s_{NN}}=$ 7.7 - 200 GeV. The global polarization and average freeze-out vorticity are also shown, and a special discussion is devoted to the polarization splitting in Au+Au 7.7 GeV collisions. Finally, a summary is drawn. Throughout this paper, we use the natural units: $\hbar = c = k_{\rm B} = 1$.



\section{Meson field in rotating system}
Considering the strong interaction of any fermions mediated by any bosonic fields, one could always write down a general equation of Lagrangian density
\be
{\cal L} =  {\cal L}_{f}+{\cal L}_{b}+{\cal L}_{int} \ .
\label{LG}
\ee 
where ${\cal L}_{f}$ denotes the Lagrangian density for the fermions, ${\cal L}_{b}$ represents the Lagrangian density for the bosons, and ${\cal L}_{int}$ is the interaction Lagrangian density  between them.
In a simplest case, this equation can be written as: 
\ba
\lefteqn{{\cal L}= \sum_i \bar{\psi}_i ( i \! \not\!\partial - m_{i} 
+ f_{\sigma} g_{\sigma} \sigma - f_{V} g_{V} \not\!V ) \psi_i} \nonumber \\
&& + \thalf \left( \partial_{\mu} \sigma \partial^{\mu} \sigma
- m_{\sigma}^2 \sigma^2 \right)
 - \oneqt V^{\mu\nu}V_{\mu\nu}
+ \thalf m_{V}^2 V_{\mu}V^{\mu} \, ,
\label{Leff}
\ea
where the first line corresponds to $({\cal L}_{f} + {\cal L}_{int})$, denoting the Lagrangian density of Dirac field for fermions with a Yuwaka interaction coupling. The second line corresponds to ${\cal L}_{b}$, being the Lagrangian density for the scalar boson $\sigma$ and vector boson $V_{\mu}$. Here, $g_{\sigma}$ is the coupling constant between fermion $\psi_i$ (of species $i$) and the scalar boson $\sigma$, and $g_{V}$ is the coupling constant between the fermion $\psi_i$  and the vector boson $V_{\mu}$.  $m_{i}$, $m_{\sigma}$ and  $m_{V}$ are respectively the mass of baryon, scalar meson and vector meson. The vector meson tensor is: $V_{\mu\nu} = \partial_\mu V_\nu - \partial_\nu  V_\mu$. The two constants, $f_{\sigma}$ and $f_{V}$ in the Yuwaka interaction term are parameters that should be determined case by case. 

In relativistic heavy ion collisions,  the hyperons are created at the chemical freeze out and then interact with other particles during the hadronic scattering phase. Given that the strong interaction of baryons (including hyperons) with other particles is mediated by a scalar meson field $\sigma$ and a vector meson field $V^{\mu}$, then with the constants $f_{\sigma}= f_{V} = 1$, and following from the Euler-Lagrange equations, one finds the equations of motion for these fields:
\ba
[ \gamma^{\mu} (i \partial_{\mu} - g_{V i} V_{\mu}) - (m_i - g_{\sigma i} \sigma)] \psi = 0 \, ,
\label{DF}\\
\partial_{\mu}\partial^{\mu} \sigma + m_{\sigma}^2 \sigma =  \sum_i g_{\sigma i} n_{si} \, , \label{KG}\\
\partial_\mu V^{\mu\nu} + m^2_{V} V^\nu = \sum_i g_{V i} J^\nu_i \, ,
\label{PE}
\ea
where  $n_{si} = \langle \bar{\psi} \psi \rangle$ is the scalar density of species $i$, and  $J^\mu_i = \langle \bar{\psi} \gamma^\mu \psi\rangle$ is the baryon current  of species $i$. These equations are actually the Dirac field equations with scalar and vector field coupling, the Klein-Gordon equation and the Proca equations. 
The detailed treatments of the above three equations has been demonstrated in Ref. \cite{CKW2019}.

For the Proca equation (\ref{PE}), analogous to Maxwell equations of massless photon field, it could be decomposed into Maxwell-Proca equations for vector mesons
\ba
\vec{\nabla}\cdot \vec{E}_V &=& \bar{g}_V \rho - m^2_{\sigma} V_0 \, , \quad
\vec{\nabla}\cdot \vec{B}_V = 0 \, , \quad 
\label{MP1}\\
\vec{\nabla}\times \vec{E} _V+ \frac{\partial \vec{B}_V}{\partial t}&=&0 \, , 
\vec{\nabla}\times \vec{B}_V -  \frac{\partial \vec{E}_V}{\partial t}=\bar{g}_V \vec{J}_{\rm B}+m_{V} \vec{V}, \nonumber \\
\label{MP2}
\ea
where $\bar{g}_V$ is the mean coupling constant of vector meson, the baryon density is $\rho_{\rm B}=\sum_i \psi_i^+ \psi_i$ and the baryon (three-)current is $\vec{J}_{\rm B}$. These are components of the baryon (four-) current $J^{\nu}_{\rm B}=(\rho_{\rm B}, \vec{J}_{\rm B})=\sum_i \bar{\psi} \gamma^{\nu} \psi$. Here the $\vec{E}_V$ \& $\vec{B}_V$ are the `electric' and `magnetic' components of the vector meson field, defined as:
\ba
E_i &\equiv & V_{i0} = \partial_i V_0 - \partial_0 V_i = (- \vec{\nabla} V_0 - \frac{\partial \vec{V}}{\partial t})_i \, , \label{BI1} \\
B_i &\equiv & -\frac12 \varepsilon_{ijk} V^{jk}=-\frac12 \epsilon_{ijk} (\partial^j V^k - \partial^k V^j) = (\vec{\nabla} \times \vec{V})_i  \, , \nonumber \\
\label{BI2}
\ea
where $i, j, k =1, 2, 3$. Let us take the curl of Maxwell-Proca equations (\ref{MP2}), and we obtain
\ba
\frac{\partial^2 \vec{E}_V}{\partial t^2} - \nabla^2 \vec{E} _V+ m_V^2 \vec{E} _V&=& 
-\bar{g}_V (\vec{\nabla} \rho_{\rm B} + \frac{\partial \vec{J}_{\rm B}}{\partial t}) , 
\label{SecondD1}\\
 \frac{\partial^2 \vec{B}_V}{\partial t^2} - \nabla^2 \vec{B}_V+ m_V^2 \vec{B}_V &=& 
 \bar{g}_V (\vec{\nabla} \times \vec{J_{\rm B}}) \, .
 \label{SecondD2}
\ea
A simple solution was obtained: 
\ba
\vec{B}_V = \frac{\bar{g}_{V}}{m_{V}^2} (\vec{\nabla} \times \vec{J}_{\rm B})\, ,
\label{SL3}
\ea
by neglecting the derivatives in eqs. (\ref{SecondD1}, \ref{SecondD2}) due to large meson mass, $m_{\omega} = 783$ MeV  and $m_{\sigma}= 550$ MeV.
Assuming global equilibrium of the system, so that $\vec{\nabla} \rho = 0$, then for the current $\vec{J}_{\rm B}=\rho_{\rm B}(\vec{x},t) \vec{v}(\vec{x},t)$, we have
\ba
\vec{\nabla} \times \vec{J}_{\rm B} =\rho_{\rm B} \,  (\vec{\nabla} \times \vec{v}) 
= \rho_{\rm B} \vec{\omega}\, , 
\label{VT}
\ea
where $\vec{\omega}=\vec{\nabla} \times \vec{v}$ is the vorticity of baryon current. Therefore, we could see that the  vortical baryon current will induce a vector meson's `magnetic' field, which, together with  its `electric' component, follow from the Maxwell-Proca equations (\ref{MP1},\ref{MP2}) and definition equations (\ref{BI1},\ref{BI2}). 

Then the non-relativistic Zeeman energy term in the Foldy-Wouthuysen (FW) Hamiltonian for the hyperon particle's spin (with effective mass $M_H$) and the vector meson's magnetic fields was written as \cite{CKW2019}:
\be
H^V_{\rm spin-B} =
- \frac{g_{\rm VH}}{M_{\rm H}} \beta \
\vec{S} \cdot \vec{B}_{\rm V} \, ,
\label{e2}
\ee
where it was argued that the constant matrix $\beta$
acting on the spin vector $\vec{S}$, will result into the opposite signs for $\Lambda$ and $\bar{\Lambda}$, thus it
might be the source of the polarization splitting.

Supposing that the spin-1/2 hyperons are in a globally equilibrated system, one could add into the density matrix of the system, $\rho$, an extra term $\rho_s \sim \exp{(\hat{\vec{S}} \cdot \Omega / T)} $, where $\Omega = \mu \vec{B}_V /S = 2 \mu \vec{B}_V$ is the vector meson's `magnetic moment' with $ \mu= -(g_{\rm VH} / M_{\rm H})\beta$ being the `magneton'.
The ensemble average of the spin vector of spin-1/2 particles are given as $\vec{S} = {\rm tr} (\rho \hat{\vec{S}})$ where $\hat{\vec{S}}$ is the spin operator. Then the ensemble averaged polarization vector in Boltzmann statistic limit can be obtained as \cite{BK2017}
\ba
\vec{P} = 2\vec{S} = \tanh \left(\frac{\Omega}{2T}\right) \hat{\vec{\Omega}}  
\simeq  \frac{\vec{\Omega}}{2T} 
= - \beta \frac{g_{\rm VH}}{M_{\rm H}} \frac{\vec{B}_{\rm V}}{T} \ ,
\label{HP}
\ea
where $\hat{\vec{\Omega} }$  is the unit vector along $\vec{\Omega}$ direction.
Taking eqs. (\ref{SL3}) and  (\ref{VT}) into the above equation, the polarization splitting would be
\ba
\Delta\vec{P} = \vec{P}_{\bar{\rm H}}- \vec{P}_{\rm H}= 2 \frac{g_{\rm VH} \bar{g}_{\rm V}}{M_{\rm H}m_{\rm V}^2} \frac{\rho_{\rm B} \vec{\omega}}{T} = C\,  \frac{\rho_{\rm B} \vec{\omega}}{T}\, ,
\label{HPD}
\ea
where $C= 2 (g_{\rm VH} \, \bar{g}_{\rm V})/(M_{\rm H}m_{\rm V}^2)$ is a coefficient determined by strong coupling constants, hyperon and meson mass.
Hence, if the baryons in high energy collisions have collectively vortical flow motion, the meson interaction with baryons can provide a mechanism for hyperon polarization splitting.

However, the equilibrium reached in high energy collision system is always assumed to be not global, but local. Thus the eq. (\ref{VT}) is actually local, and should be modified:
\ba
\vec{\nabla} \times \vec{J}_{\rm B}
= \rho_{\rm B}\, \vec{\omega} + \vec{\nabla}\rho_{\rm B} \times \vec{v} \, , 
\label{CC}
\ea
and the average polarization splitting eq. (\ref{HPD}) becomes:
\ba
\Delta \vec{P}_{J} = 
\langle C\frac{\vec{\nabla} \times \vec{J}_{\rm B}}{T} \rangle
&&= C\langle\frac{\rho_{\rm B} \, \vec{\omega}}{T} \rangle
+ C\langle\frac{ \vec{\nabla}\rho_{\rm B} \times \vec{v} }{T} \rangle\, \nonumber \\
&&=\Delta\vec{P}_{\omega}+\Delta\vec{P}_{\rho}.
\label{HPP}
\ea
where $\langle ... \rangle$ denotes the average over the space. Here $\Delta \vec{P}_J$ is the average polarization splitting induced by rotating baryon current $\vec{J}_{\rm B}$, $\Delta\vec{P}_{\omega}=\langle C(\rho_{\rm B} \, \vec{\omega})/T \rangle$ originates from the vorticity $\omega$ only, and $\Delta\vec{P}_{\rho}=\langle C(\vec{\nabla}\rho_{\rm B} \times \vec{v})/T \rangle $ results from the baryon density gradient. 

In this work, the values of coefficients in eq. (\ref{HP}) are kept the same as in Ref. \cite{CKW2019}: $M_{\Lambda} = 1115.6$ MeV, $M_{\rm V} = 780$ MeV, $\bar{g}_{\rm V} = 5$, and $g_{\rm V\Lambda} \approx 0.55 g_{\rm VN} \approx  4.76$. Besides, noting that the Foldy-Wouthuysen transformation used to deduce eq. (\ref{e2}) is non-relativistic and so are the ensuing eqs. (\ref{HP},\ref{HPD},\ref{HPP}), thus we assume that the post-freeze-out system is near to the Boltzmann limit, and the $\Lambda$ particles are non-relativistic. Then to compared with the experimental results, $\Delta P$ in equations (\ref{HP},\ref{HPD},\ref{HPP}) 
should be Lorentz-boosted from pre-freeze-out center-of-mass frame into the $\Lambda$'s rest frame, just like the polarization 3-vector $\vec{\Pi}( p )$ is Lorentz-boosted into the particle's rest frame via: 
\be
\vec{\Pi}_0(\vec{p})=
\vec{\Pi}( p )-\frac
{\vec{p}}
{p^0 (p^0 + m)}
\vec{\Pi}( p ) \cdot \vec{p} \ .
\label{Pi0}
\ee
However, according to our calculations, the polarization in $\Lambda$ frame is only $\sim$0.3\% smaller than that in QGP frame, or corresponding to only 5-10\% correction (especially at the low energies 7.7 - 30 GeV). This is because most of the particles dwell in the low transverse momentum space, and thus the boost effect is also small. Thus we believe that the present calculations based on eqs. (\ref{HP}) or (\ref{HPP}) are satisfactory quantitative estimates.

\section{Gloabl polarization and its splitting at $\sqrt{s_{NN}}= 7.7- 200$ GeV }

\begin{figure}[h] 
\begin{center}
\includegraphics[width=8.0cm]{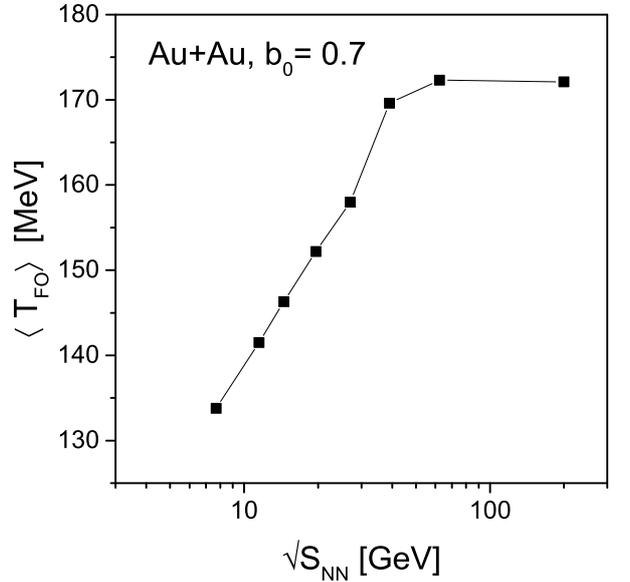}
\end{center}
\vspace{-0.3cm}
\caption{
(Color online) The averaged freeze-out temperature $\langle T_{\rm FO} \rangle$ at different collision energy with freeze-out time varying among 5.9 - 7.9 fm/c.
}
\label{F1}
\end{figure}

The nucleus-nucleus impact in our initial state is divided into many slab-slab collisions, and Yang-Mills flux-tubes. These are assumed to form  streaks
\cite{M1,M2}. 
In this scenario, the initial state naturally generates longitudinal velocity shear flow, which when placed into the subsequent high resolution (3+1)D PICR hydrodynamic model, will develop into substantial vorticity. Since our initial state+hydrodynamic model describes the shear and vorticity in heavy ion collisions fairly well, its simulations to the $\Lambda$ polarization also achieved success.

\begin{figure}[ht] 
\begin{center}
\includegraphics[width=8.0cm]{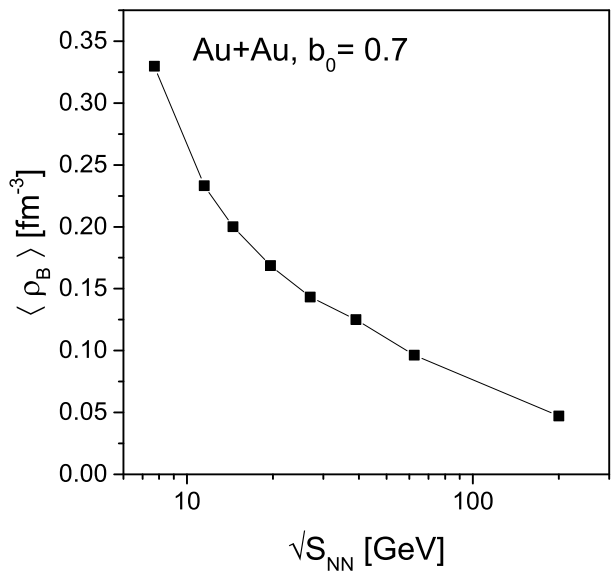}
\end{center}
\vspace{-0.3cm}
\caption{
(Color online)The averaged baryon density at freeze-out for Au+Au collisions with different collision energy.
}
\label{F2}
\end{figure}

Therefore, we use the PICR hydrodynamic model to simulate the Au+Au collisions at RHIC BES energy region $\sqrt{s_{\rm NN}}= 7.7- 200$ GeV, and calculate the global polarization with approaches developed in \cite{BCW2013}, as well as its difference between the $\bar{\Lambda}$ and $\Lambda$ based on eq. (\ref{HPP}). 

For the purpose of continuity, we do not perform a new simulation, but just use the same data in our previous Rapid Communication \cite{Xie2017}, which was then the first work to show the energy dependence of global polarization $\vec{P}_{\rm H}$, and seemed to exhibit fairly good agreement with the experimental data. In that work, the simulation parameters were set as follows: the impact parameter ratio was: $b_0=b/b_{max}= 0.7$, (where $b$ is the impact parameter and $b_{max}$ is the maximum impact parameter); the cell size was $0.343^3$ fm$^3$, the time increment is 0.0423 fm/c; the freeze-out time was fixed to be 7.24 fm/c = 2.5+4.74 fm/c for all collisions energies (2.5 fm/c for the initial state's stopping time and 4.74 fm/c corresponds to the hydro-evolution time). However, a fixed freeze-out time for different energies is actually not very physical, thus in this work, we are going to vary the freeze-out time for varied collision energies. More specifically, the freeze-out time increases from 5.9 fm/c to 7.9 fm/c with the collision energy increasing from 7.7 GeV to 200 GeV, so that the average temperature of the system at freeze-out, as shown in Fig. \ref{F1}, agrees with the theoretical calculations and experimental results\cite{BMG2006,STAR2013}. The average baryon densities at the chosen freeze-out time are also shown in Fig. \ref{F2}, whose values are at the same scale of the freeze-out charge densities in AMPT model as shown in Fig. 2 of Ref. \cite{GLW2019}.

\begin{figure}[ht] 
\begin{center}
      \includegraphics[width=8.5cm]{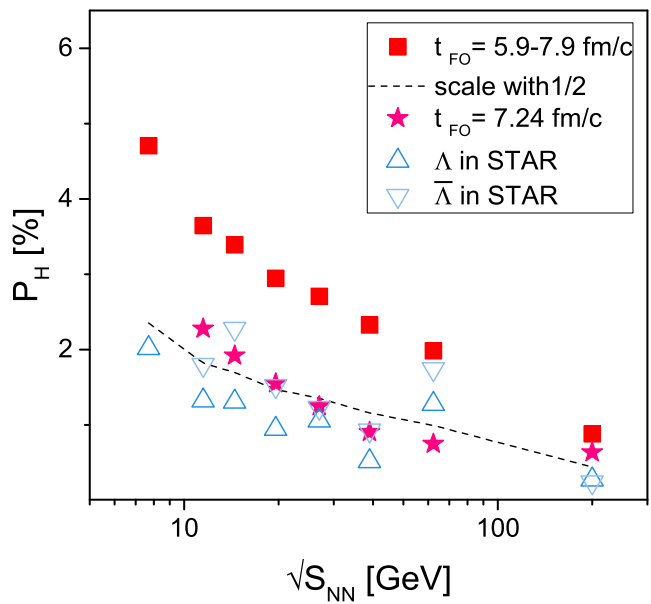}
\end{center}
\caption{
(Color online) The global $\Lambda$ polarization for Au+Au collisions at $\sqrt{s_{NN}}=7.7-200$ GeV with impact parameter ratio $b_0=0.7$. The red stars are extracted from our previous work\cite{Xie2017},  showing the global polarization as a function of collision energy, with the freeze-out time being fixed to $t_{\rm FO}= 7.24$ fm/c. The red squares correspond to the case of varied freeze-out time, $t_{\rm FO}$= 5.9 - 7.9 fm/c, with varied collision energy $\sqrt{s}$= 7.7 - 200 GeV. The experimental data denoted by up or down triangles are extracted from Ref. \cite{QMSTAR}. 
}
\label{F3}
\end{figure}

The red squares in Fig. \ref{F3} show the global $\Lambda$ polarization in our model for Au+Au collisions at $\sqrt{s_{NN}}=7.7-200$ GeV, with varied freeze-out time $t_{\rm FO}= 5.9 - 7.9$ fm/c. As comparison, we also show  in Fig. \ref{F3} our previous results with fixed $t_{\rm FO}= 7.24$ fm/c \cite{Xie2017} , by pink stars. One can see that the new values of global polarization is larger than old ones, showing the sensitivity of global polarization to the freeze-out time, while the the energy dependence behavior is still kept. The large magnitude of new results are reasonable, since the collision is peripheral with centrality of $c=b_0^2=49\%$. The STAR experiment results show that the average global polarization linearly dependent on the centrality, and its value at 20\%-50\% centrality bin is about half of that at 50\% centrality\cite{QMSTAR}. Assuming that the linear dependency of global polarization on centrality is similar for different RHIC BES energies, we estimate the global polarization at 20\%-50\% centrality bin by scaling down the global polarization at $b_0=0.7$ with factor of 0.5, and show them with the dashed line in Fig. \ref{F3}. One could see the estimated values are very close to the experimental results which are denoted by the up or down triangles. Besides, the correction from the feed-down effect of resonance decay would turn down the values by about 15\%-20\% \cite{XLHH2019,BCS2019}.

\begin{figure}[ht] 
\begin{center}
      \includegraphics[width=8.5cm]{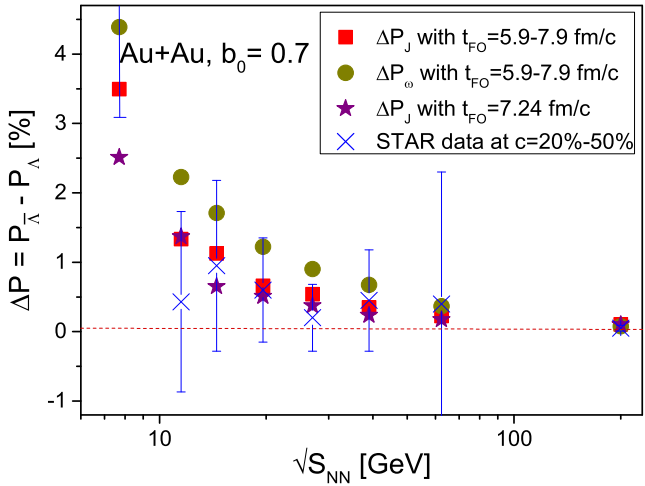}
\end{center}
\caption{
(Color online) The polarization splitting between $\Lambda$s and $\bar{\Lambda}$s for Au+Au collisions at $\sqrt{s_{NN}}=7.7-200$ GeV with impact parameter ratio $b_0=0.7$. The red squares represent the polarization splitting $\Delta P_J$ in eq. (\ref{HP}), with varied freeze-out time $t_{\rm FO}= 7.24$ fm/c. The green circles correspond to polarization splitting $P_{\omega}$ which is induced by vorticity only. As comparison, $\Delta P_J$ with fixed $t_{\rm FO}= 7.24$ fm/c are shown by purple stars. The experimental data denoted by cross symbols with error bars are extracted from Ref. \cite{QMSTAR}, as an average for centralities $c = $ 20-50\%, corresponding to impact parameter ratio $b_0$ = 0.45 - 0.7 \cite{XWC2019}.
}
\label{F4}
\end{figure}

\begin{figure}[ht] 
\begin{center}
\includegraphics[width=8.5cm]{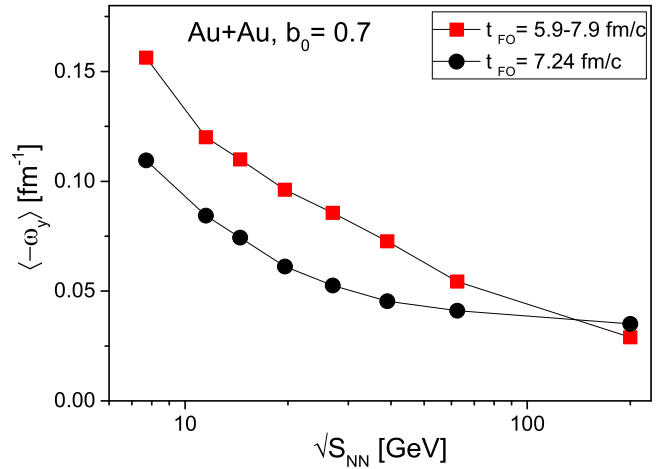}
\end{center}
\caption{
(Color online)The average $y$-directed vorticity $\langle - \omega_y \rangle$ at freeze out as function of collision energy with impact parameter ratio $b_0=0.7$. The black and red symbols respectively represent the vorticity at fixed freeze-out time $t_{\rm FO}= 7.24$ fm/c, and at varied freeze-out time $t_{\rm FO}= 5.9 - 7.9$ fm/c.
}
\label{F5}
\end{figure}

The red squares in Fig. \ref{F4} show the average polarization splitting $\Delta P_J$ of eq. (\ref{HP}) in our PICR hydrodynamic model for different  RHIC BES collision energies, with varied freeze-out time. One can see that our calculation results has the same tendency with the STAR data, which are denoted by cross symbols and error bars, and the magnitudes are also in line with each other if we ignore the different centrality settings ($b_0=0.7$ or $c \approx $  50\% for our calculation and $c$ = 20\% - 50\% for STAR). Besides, our results are similar to the Fig. 1(b) in Ref. \cite{CKW2019}, which was obtained with the free parameter $C$ assumed to be dependent on collision energy. Actually the parameter $C$ in Ref. \cite{CKW2019} should be dependent on collision energy, since $C$ is related to the system's vorticity at freeze-out, and as shown by Fig. \ref{F5}, the vorticity decreases with increasing energy. As comparison, we also show, by purple stars in Fig. \ref{F4}, the $\Delta P_J$ with fixed freeze-out time, and they are usually a bit smaller than that with varied freeze-out time.

The green circles in Fig. \ref{F4} denotes the polarization splitting $\Delta P_{\omega}$ induced by vorticity only. We can see that  $\Delta P_{\omega} > \Delta P_J$, which means the second term in eq. (\ref{HP}) induced by baryon density gradient, $\Delta P_{\rho}$, is actually negative and non-trivial. For different collision energies, the $\Delta P_{\rho}$ would downplay the final splitting effect for about $1/3 \sim 1/4$.

Now we can compare the splitting effect denoted by red squares in Fig. \ref{F4}, to the calculated polarization denoted by red squares in Fig. \ref{F3}. One can see that the splitting effect induced by the meson field leads to a limited correction to the global polarization. Taking the case of 11.5 GeV for example: the splitting effect $\Delta P_J$ with varied FZ time is about 1.33\%, and the corresponding polarization is about 3.65\%. Thus the correction for the $\Lambda$ polarization is $1.33\%/2/3.65\% \approx 18\%$. Typically, the corrections for the global polarization for different collision energies are about 5\%- 20\%, except for 7.7 GeV (about 30\%).

Fig. \ref{F5} shows the averaged vorticity along the $y$ direction over the overlapping region
\ba
\langle \omega_y \rangle = \langle  [\vec{\nabla} \times  \vec{v} ]_y  \rangle  \, ,
\ea
 as a function of collision energy. It is not  surprising  that the $y$-directed vorticity decreases with the collision energy, and the magnitude of the vorticity in our model is the same scale of that from the AMPT simulation. 
We take the case of $\sqrt{s_{\rm NN}}$ = 200 GeV for example. In our model the vorticity $\omega_y$ at freeze-out for $\sqrt{s_{\rm NN}}$ = 200 GeV with $b_0$=0.7 (or b=9.6fm), is about 0.25 - 0.35. Meanwhile the AMPT model shows, with $b=9$ fm, the vorticity value at late time (Fig. 11 in ref. \cite{JLL2016}) is around 0.2 - 0.3, and the thermal vorticity (Fig. 1 in ref. \cite{WDH2019}) is about  0.15.
\footnote{To compare vorticity herein with the thermal vorticity
$\vec{\varpi} =  \frac12 \vec{\nabla} \times ( \gamma  \vec{v} / T )$
defined in Ref. \cite{WDH2019}, one could estimate the freeze-out temperature as around 170 MeV at $\sqrt{s}=39-200$ GeV \cite{BMG2006}, and thus the factor $\frac12\frac{\hbar}{T} \approx 0.51 $ fm/c.}

Finally we want to have a little discussion on the polarization splitting at 7.7 GeV. For the case of 7.7 GeV, the polarization difference from our model could be as significant as $\Delta P_J \approx $ 3.5\%, which is already larger than the lower boundary of experimental measurement of 3\%. Up to now several mechanisms were proposed, and quantitative calculations were performed to explain the $\Lambda$ and $\bar{\Lambda}$ polarization splitting
\cite{GLW2019,CKW2019,HX2018,VBZ2019}, 
but none of them can achieve 3\% difference at 7.7 GeV. More specifically, our result for 7.7 GeV case is about 3 times larger than the upper boundary estimate in Ref. \cite{GLW2019}. As discussed before that the values of density quantity and the vorticity between our model and the AMPT model are very close, the reason why we have much larger polarization splitting effect than that in Ref. \cite{GLW2019} lies on the coefficient $C= 2 (g_{\rm VH} \bar{g}_{\rm V})/(M_{\rm H}m_{\rm V}^2)$, which contains strong coupling constants that are much `stronger' than the weak coupling constants in Ref. \cite{GLW2019}.

\begin{figure}[ht] 
\begin{center}
\includegraphics[width=8.5cm]{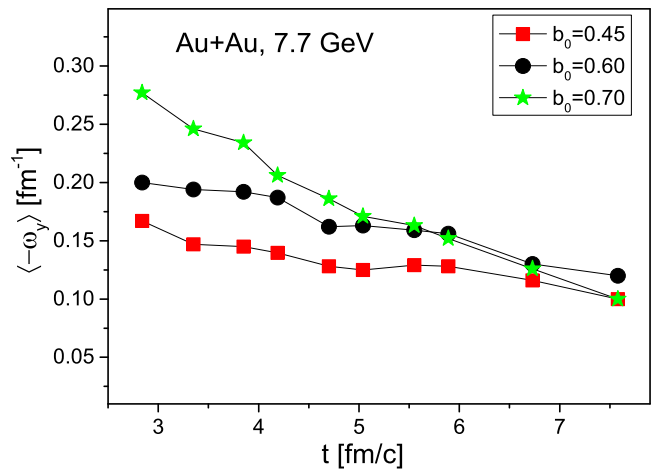}
\end{center}
\caption{
(Color online)The time evolution of average $y$-directed vorticity $\langle - \omega_y\rangle$ with different impact parameters in our model for Au+Au collisions at 7.7 GeV.
}
\label{F6}
\end{figure}

One might argue that in the more central collisions of $b_0 < 0.7$, the vorticity will decrease and then the overall polarization splitting at 7.7 GeV would be suppressed to lower than 3\%. To deal with this issue, we show the Table \ref{tab1}, where we calculate the polarization splitting $\Delta P_J $ at different centralities, for Au+Au 7.7 GeV collisions. Surprisingly, within the centrality range of 20-50\%, the polarization splitting $\Delta P_J$ is actually larger for more central collisions. Two factors lead to this unexpected effect:

(1) Note that the freeze-out condition herein is a constant temperature for the whole centrality range 20\%-50\%, and as one can see in Table \ref{tab1},  this leads to a smaller freeze-out time for more central collisions than that for peripheral collisions.
Meanwhile, Fig. \ref{F6} shows that the average vorticity at Au+Au 7.7 GeV collisions in our model has a very mild decreasing tendency with the evolution time (similar behavior was seen by UrQMD model at 2 GeV\cite{DHMZ2020}). Therefore, the vorticity at freeze out for different centralities are rather close, which results into a similar value of $\Delta P_{\omega}$ in Table \ref{tab1} for different centralities.

 (2) Then, the larger fluctuation of baryon density in peripheral collisions means a larger $|\Delta P_{\rho}|$, which of course results into a smaller polarization splitting $\Delta P_J$ in peripheral collisions. 
 
Therefore it indicates that for Au+Au 7.7 GeV collisions at the centrality range of 20\%-50\%, the polarization splitting will be larger than 3.5\%. Furthermore, we have checked that at  the energy of $\sqrt{s}_{NN}\ge$ 11.5 GeV the above effect no longer exists, because the freeze-out time is larger for more central collisions, and then the vorticity as well as the polarization splitting is smaller for more central collisions. Thus for  the energy of $\sqrt{s}_{NN}\ge$ 11.5 GeV, the polarization splitting at 20\%-50\% centrality bin will be suppressed compared to that at $b_0=$ 0.7( or $c$=49\%) shown in Fig. \ref{F4}. One can check this from Table \ref{tab2} for the case of 11.5 GeV. The key point here is that the dependency of the freeze-out time on the centrality at 7.7 GeV is opposite to that at 11.5 GeV and beyond, and we are presently not sure the mechanism behind this. It is possible that our initial state + PICR model loses its validity for low energy collisions of 7.7 GeV, or it might imply the fluid dynamics is different in the collision system of 7.7 GeV and 11.5 GeV. This effect might be exactly the reason why the signal of polarization splitting observed at STAR Au+Au 7.7 GeV collisions seems so strong compared to that at other collision energies, but after all it needs more investigations and confirmations from other models.


\renewcommand\arraystretch{1.5}
\begin{table}
\begin{center}
\setlength{\tabcolsep}{2.5mm}{
     \begin{tabular}{ l | l | l | l }
        \hline\hline
            $b_0$ (c) &0.45 (20\%)&0.6 (36\%)&0.7 (49\%) \\ \hline
            $\langle T_{\rm FO}\rangle$ (MeV) &134.3 & 134.8 & 133.8 \\ 
            $t_{\rm FO}$   (fm/c) &4.2 &5.1 & 5.9  \\
            $\langle \rho_{\rm B} \rangle$   (fm$^{-3}$) &0.36 & 0.345 &0.33\\ 
            $\langle - \omega_y \rangle$   (fm$^{-1}$) &0.140 & 0.163 &0.156\\ 
            $\Delta P_{\omega}$ &4.49\% & 4.77\%&4.39\%\\
             $\Delta P_{J}$ &4.28\% & 4.12\%&3.49\%\\
         \hline\hline
      \end{tabular}}
\end{center}
\caption{The average freeze-out temperature $T_{\rm FO}$, freeze-out time $t_{\rm FO}$, average baryon density $\langle \rho_{\rm B} \rangle$, average vorticity $\langle - \omega_y \rangle$, and $\Delta P_{\omega}$, $\Delta P_{J}$ defined in eq. (\ref{HP}), for Au+Au 7.7 GeV collisions at different centalities $c=$ 20\%, 36\%, 49\% .}
\label{tab1}
\end{table}

\begin{table}
\begin{center}
\setlength{\tabcolsep}{2.5mm}{
     \begin{tabular}{ l | l | l }
        \hline\hline
            $b_0$ (c) &0.45 (20\%)&0.7 (49\%) \\ \hline
            $\langle T_{\rm FO}\rangle$ (MeV) &142.2 & 141.5\\ 
            $t_{\rm FO}$   (fm/c) &7.9& 5.9  \\
            $\langle - \omega_y \rangle$  (fm$^{-1}$) &0.095 &0.120\\ 
            $\Delta P_{\omega}$ &1.43\% &2.23\%\\
             $\Delta P_{J}$ &0.83\% &1.33\%\\
         \hline\hline
      \end{tabular}}
\end{center}
\caption{The average freeze-out temperature $T_{\rm FO}$, freeze-out time $t_{\rm FO}$, average vorticity $\langle - \omega_y \rangle$, and $\Delta P_{\omega}$, $\Delta P_{J}$ defined in eq. (\ref{HP}), for Au+Au 11.5 GeV collisions at different centalities $c=$ 20\%, 49\% .}
\label{tab2}
\end{table}

\section{Summary and Conclusion}
With the PICR hydrodynamic model, we study the polarization splitting between $\Lambda$ and $\bar{\Lambda}$ , based on the meson field mechanism. Our results fit to the experimental data fairly well. 
Two unexpected effect emerges: (1) the baryon density gradient has non-trivial and negative contribution to the polarization splitting; (2) for 7.7 GeV Au+Au collisions within the centrality range of 20\%-50\%, the polarization splitting surprisingly increases with the centrality decreases. The second effect might hint  different hydrodynamics for collision system with energy below 7.7 GeV, and help to explain the significant signal of polarization splitting measured in STAR's Au+Au 7.7 Gev collisions. The reason why the Au+Au collisions  at 7.7 GeV in our model has a longer evolution time for peripheral collisions, is still unclear and needs more investigations.

\section*{Acknowledgments}

We thank  Dujuan Wang for enlightened discussions. 
The work of of Y. L. Xie is supported by the National Natural Science Foundation of China(12005196), and the Fundamental Research Funds for the Central Universities (G1323519234), of L. P. Cs. is supported by the Research Council of Norway, and of G. Chen is supported by National Natural Science Foundation of China(11475149).


\begin{thebibliography}{}
\bibitem{BPR2008}
 F. Becattini, F. Piccinini, and J. Rizzo,
 Phys. Rev. C {\bf 77}, 024906 (2008).

\bibitem{GCD2008}
 J.-H. Gao, S.-W. Chen, W.-T. Deng, and Z.-T. Liang, Q. Wang, and X.-N. Wang, 
 Phys. Rev. C {\bf 77}, 044902 (2008).

 \bibitem{VAC2014}
V. Vovchenko, D. Anchishkin, L.P. Csernai,
Phys. Rev. C {\bf 90}, 044907 (2014)
 
 \bibitem{EH1915}
A. Einstein and W. de Haas, Deutsche Physikalische Gesellschaft, Verhandlungen {\bf 17}, 152 (1915).

\bibitem{B1935}
S. J. Barnett, Rev. Mod. Phys. {\bf 7}, 129 (1935).

\bibitem{BCDG13}
 F. Becattini, V. Chandra, L. Del Zanna, and E. Grossi,
 Annals of Phys. {\bf 338}, 32 (2013).

\bibitem{BCW2013}
F. Becattini, L.P. Csernai, D.J. Wang,
Phys. Rev. C {\bf 88}, 034905 (2013).
 
 \bibitem{IA06}
 I. Abt {\it et al.} (HERA-B Collaboration), 
 Phys. Lett. B {\bf 638}, 415-421, (2006).
 
\bibitem{IS06}
 I. Selyuzhenkov {\it et al.} (STAR Collaboration),
 J. Phys. G: Nucl. Part. Phys. {\bf 32}, S557-S561 (2006).

\bibitem{BIA07}
 B. I. Abelev {\it et al.},
 Phys. Rev. C {\bf 76}, 024915 (2007).

 \bibitem{LW05a}
 Z.-T. Liang, and X.-N. Wang,
 Phys. Rev. Lett. {\bf 94}, 102301 (2005).
 
\bibitem{HHW11}
 X.-G. Huang, P. Huovinen, and X.-N. Wang,
 Phys. Rev. C {\bf 84}, 054910 (2011).
 
\bibitem{BGT07}
 B. Betz, M. Gyulassy, and G. Torrieri,
 Phys. Rev. C {\bf 76}, 044901 (2007).

\bibitem{FirstSTAR}
B. I. Abelev {\it et al.} (STAR Collaboration),  Phys. Rev. C {\bf 76}, 024915 (2007), [Erratum: Phys. Rev. C {\bf 95}, 039906 (2017)].

\bibitem{Nature}
L. Adamczyk {\it et al.} (The STAR Collaboration), Nature {\bf 548}, 62 (2017).

\bibitem{SecondSTAR}
J. Adam {\it et al.} (STAR Collaboration), Phys. Rev. C {\bf 98}, 014910 (2018).

\bibitem{QMSTAR}
T. Niida {\it et al}. (STAR Collaboration), Invited talk at Quark Matter 2018, May 13-19, 2018, Venice, Italy.

\bibitem{XLTW2018}
Xiao-Liang Xia, Hui Li, Zebo Tang, and Qun Wang, Phys. Rev. C {\bf 98}, 024905 (2018).

\bibitem{BK2018}
F. Becattini and I. Karpenko, Phys. Rev. Lett. {\bf 120}, 012302 (2018).

\bibitem{Xie2016}
Y. L. Xie, M. Bleicher, H. St\"ocker, D. J. Wang, and L. P. Csernai, Phys. Rev. C {\bf 94}, 054907 (2016).

\bibitem{STAR2019}
J. Adam {\it et al.} (STAR Collaboration), Phys. Rev. Lett. {\bf 123}, 132301 (2019).


\bibitem{SK2019}
Y. F. Sun and C. M. K, Phys. Rev. C {\bf 99}, 011903(R) 2019.

\bibitem{WPHW2019}
H.-Z. Wu, L.-G. Pang, X.-G. Huang, and Q. Wang, Phys. Rev. Research {\bf 1}, 033058 (2019).

\bibitem{FKMR2019}
W. Florkowski, A. Kumar, A. Mazeliauskas, and R. Ryblewski, Phys. Rev. C {\bf 100}, 054907 (2019).

\bibitem{XLHH2019}
X.-L. Xia, H. Li, X.-G. Huang, and H. Z. Huang, Phys. Rev. C {\bf 100}, 014913 (2019).

\bibitem{BCS2019}
F. Becattini, G. Q Cao, E. Speranza, Eur. Phys. J. C {\bf 79}, 741 (2019).

\bibitem{XWC2019}
Yilong Xie, Dujuan Wang, Laszlo P. Csernai, Eur. Phys. J. C {\bf 80},39 (2020).

\bibitem{GLW2019}
X. Y. Guo, J. F. Liao, E. K. Wang, Scientific Reports {\bf 10}, 2196 (2020).

\bibitem{CKW2019}
L. P. Csernai, J. I. Kapusta, and T. Welle, Phys. Rev. C {\bf 99}, 021901(R) (2019).

\bibitem{BK2017}
F. Becattini {\it et al}., Phys. Rev. C {\bf 95}, 054902 (2017).

\bibitem{M1}
V. K. Magas, L. P. Csernai, and D. D. Strottman,
 Phys. Rev. C {\bf 64} 014901 (2001).

\bibitem{M2}
V. K. Magas, L. P. Csernai, and D. D. Strottman,
 Nucl. Phys. A {\bf 712} 167 (2002).

\bibitem{Xie2017}
Y. L. Xie, D. J. Wang, L. P. Csernai, Phys. Rev. C {\bf 95}, 031901R (2017).

\bibitem{BMG2006}
F. Becattini, J. Manninen, M. Gazdzicki, Phys. Rev. C {\bf 73}, 044905 (2006)

\bibitem{STAR2013}
S. Das {\it et al.} (STAR Collaboration), Nucl. Phys. A {\bf 904-905}, 891c (2013).

\bibitem{JLL2016}
Y. Jiang, Z.-W. Lin, and J. F. Liao, Phys. Rev. C {\bf 94}, 044910 (2016).

\bibitem{WDH2019}
D.-X. Wei, W.-T. Deng, and X.-G. Huang, Phys. Rev. C {\bf 99}, 014905 (2019).

\bibitem{HX2018}
Z.-Z. Han, J. Xu, Phys. Lett. B {\bf 786}, 255 (2018).

\bibitem{VBZ2019}
O. Vitiuk, L. Bravina, E. Zabrodin, Phys. Lett. B {\bf 803}, 135298 (2020).

\bibitem{DHMZ2020}
X.-G. Deng, X.-G. Huang, Y.-G. Ma and S. Zhang, Phys. Rev. C {\bf 101}, 064908 (2020).


\end{thebibliography}
\end{document}